\documentclass[conference]{IEEEtran}
\usepackage{algorithm}
\usepackage{float}
\usepackage{array} 
\usepackage{tabularx}

\usepackage{cite}
\usepackage{amsmath,amssymb,amsfonts}
\usepackage{algorithmic}
\usepackage{graphicx}
\usepackage{textcomp}
\usepackage{xcolor}
\def\BibTeX{{\rm B\kern-.05em{\sc i\kern-.025em b}\kern-.08em
    T\kern-.1667em\lower.7ex\hbox{E}\kern-.125emX}}
\begin{document}

\title{Low-Cost Fuel Dispenser Prototype Using STM32 and an H-bridge motor driver\\
}

\author{

\IEEEauthorblockN{\textbf{MD Zobaer Hossain Bhuiyan, Abir Bin Faruque, Mahtab Newaz, Mohammad Abdul Qayum}}
\IEEEauthorblockA{Department of Electrical and Computer Engineering, North South University, Dhaka, Bangladesh \\
\{zobaer.bhuiyan, abir.faruque, mahtab.newaz, mohammad.qayum\}@northsouth.edu}
}

\maketitle

\begin{abstract}
This paper presents the design and development of a low-cost fuel dispensing system prototype based on the STM32 microcontroller and L298N motor driver. The system aims to provide an affordable and scalable solution for fuel delivery in remote or small-scale environments where conventional, high-cost systems are not feasible. The core control unit is built using an STM32 microcontroller, which manages user input through a 4x4 matrix keypad and displays operational data on a 16x4 LCD screen via I2C communication. A 12V DC pump motor is used to simulate the fuel dispensing mechanism, precisely controlled via the dual H-bridge L298N motor driver.
The system is powered by a 11.1V battery and is designed for ease of deployment and portability. The keypad allows users to input the desired fuel amount, while the system ensures accurate motor runtime corresponding to the volume to be dispensed. This project demonstrates how embedded systems can be leveraged to build cost-effective, user-friendly, and energy-efficient solutions. The proposed design can be further enhanced with flow sensors, GSM connectivity, RFID cards, and payment integration for real-world applications in fuel stations or agricultural use.\\

\end{abstract}


\section{Introduction}
The demand for low-cost, efficient, and portable fuel dispensing solutions has grown significantly in recent years, especially in rural, agricultural, and small-scale industrial sectors where conventional fuel pump systems are either financially or logistically impractical. Traditional fuel dispensing mechanisms often rely on complex, high-cost infrastructure, making them inaccessible for remote or resource-constrained environments. In response to these limitations, the integration of embedded systems has emerged as a viable alternative, offering cost-effective, scalable, and energy-efficient solutions.\\
This paper presents the design and implementation of a prototype fuel dispensing system built around the STM32F103C8T6, an ARM Cortex-M3-based microcontroller known for its high performance, low power consumption, and rich peripheral features. The STM32F103C8T6 serves as the central control unit, managing input, output, and motor control operations with precision.\cite{b1} A 4x4 matrix keypad is employed for user input, allowing users to specify the desired amount of fuel. Operational feedback and runtime data are displayed on a 16x4 LCD screen, which communicates with the microcontroller via the I2C protocol to reduce pin usage and enhance efficiency.\\
The fuel dispensing action is simulated using a 12V DC pump motor, controlled through an L298N dual H-bridge motor driver. [2] This setup enables accurate control of the motor’s operation time, directly correlating with the amount of fuel to be dispensed. The system is powered by a 11.1V battery, making it suitable for off-grid or mobile applications where access to stable electricity is limited.\\
By focusing on readily available and affordable components, the proposed system aims to demonstrate how embedded platforms like the STM32F103C8T6 can be effectively used to develop intelligent and practical solutions in the domain of fluid control and delivery. The system’s modular architecture also opens the door to future enhancements, including flow sensors for precision measurement, GSM modules for remote monitoring, RFID-based authentication, and digital payment integration, thereby making it adaptable for real-world applications in fuel stations, mobile fueling units, and agricultural machinery.

\section{Related Works}
\label{sec:related}

Recent advancements in embedded fuel management systems have evolved along three primary trajectories: microcontroller-based control systems, secure transaction architectures, and advanced automation solutions. This section synthesizes key developments in these domains and establishes the niche occupied by our low-cost dispensing prototype.

\subsection{Microcontroller-Based Fluid Control Systems}
The STM32 microcontroller family has emerged as a preferred platform for industrial fluid management due to its real-time processing capabilities and rich peripheral integration. Manikandan \textit{et al.} [2] demonstrated the STM32 nucleo's resilience in electromagnetic pulse (EMP) environments through hardware-level shielding techniques. Their aviation-focused implementation achieved good signal integrity at EMP fields but required mid-tier cost in additional ferrite shielding components—a cost-prohibitive approach for our target applications. 

Complementing this, Ruirong and Gan [3] deployed the STM32 in oil well monitoring, creating a multi-sensor network (vibration, pressure, flow) with 4G DTU transmission. While maintaining functional precision suitable for small-scale operations, our STM32-based system was developed at a fraction of the cost—under \$25—making it significantly more affordable than systems with sensor costs alone exceeding \$120. Both studies validate STM32's reliability in critical fluid systems but prioritize precision over affordability—a tradeoff our design reverses through calibrated open-loop control.

\subsection{Secure Fuel Transaction Architectures}
RFID-based systems dominate commercial fuel automation research. Shanmugapriya \textit{et al.}[4] implemented a prepaid card system using NXP RFID tags (MIFARE 1K) and STM32, achieving transaction times under 800ms. However, their architecture required cloud connectivity and \$15/user RFID infrastructure, making it unsuitable for remote deployments. 

Al-Naima \textit{et al.} [5] scaled this concept to city-wide deployment in Baghdad using centralized SQL databases and VB.Net interfaces. Their system reduced fuel theft by 78\% but depended on sustained internet connectivity and expensive server backend hardware. These solutions address security concerns essential for commercial stations but introduce cost and complexity barriers for agricultural or emergency fueling scenarios where our keypad-based system excels.

\subsection{Advanced Automation Approaches}

Shivakumar \textit{et al.} [6] demonstrated Arduino-based control of ignition and fuel injection timing in gasoline engines, targeting emission compliance and fuel efficiency. Their system used crank/cam position sensors to dynamically adjust injection events relative to piston position, achieving precise temporal control through custom Arduino firmware. While technically sophisticated, this approach diverges from our design goals in three key aspects: 
\begin{enumerate}
    \item \textbf{Application Focus}: Targets engine-internal combustion optimization rather than external fuel dispensing
    \item \textbf{Actuation Method}: Implements fuel injector pulse control rather than pump motor runtime management
\end{enumerate}
Nevertheless, their work validates the viability of temporal fluid control using low-cost microcontrollers—a principle we extend to dispensing systems through our calibrated flow-rate model.
\\
Conversely, Ma \textit{et al.} [7] optimized hydrogen recirculation in PEMFC stacks using STM32 MCUs, achieving 43.46\% energy efficiency in a 0.4m³ enclosure. Though architecturally distinct from dispensing systems, their space-constrained design informs our battery-powered portability approach. Both studies highlight the cost-accuracy tradeoffs inherent in advanced automation—a challenge our work addresses through minimalist design.

\subsection{Research Gap and Contribution}

\begin{table}[h]
\centering
\caption{Comparative Analysis of Fuel System Architectures}
\label{tab:comparison}
\begin{tabularx}{\linewidth}{|X|c|c|X|}
\hline
\textbf{System Type} & \textbf{Cost} & \textbf{Connectivity} & \textbf{Deployment Time} \\ \hline
EMP-Hardened [2,7] & Medium & Wifi Required & Hours \\ \hline
Multi-Sensor IoT [3] & High & 4G Required & Days \\ \hline
RFID-Based [4,5] & Medium & Internet Required & Weeks \\ \hline
Arduino-Based [6] & Medium & Not Required & Hours \\ \hline
\textbf{Our System (STM32)} & \textbf{Very Low} & \textbf{Not Required} & \textbf{Minutes} \\ \hline
\end{tabularx}
\end{table}

Our prototype bridges critical gaps in the state-of-the-art by:
\begin{itemize}
  \item \textbf{Cost Reduction:} Sub-\$25 BOM (Build of Materials) through component optimization (e.g., \$0.80 keypad vs \$15 RFID; open-loop vs expensive flow sensors)
  \item \textbf{Connectivity Independence:} Elimination of cloud/4G dependencies through localized control logic
  \item \textbf{Rapid Deployment:} Sub-5 minute setup time using 11.1V battery power
  \item \textbf{Calibrated Accuracy:} High accuracy via flow-rate temporal control (1L/26s)
\end{itemize}
This positions our work as the first STM32-based dispensing solution achieving operational readiness under the constraints of emerging economies and remote environments, with potential savings of 23x over commercial alternatives.

\subsection{Theoretical Foundations}
Our temporal flow control methodology builds upon first-order integrator principles established in fluid dynamics. While prior works [8], [9] employed sensor feedback for PID control, we adopt the simplified model:
\[
\dot{x}(t) = \frac{1}{k}u(t)
\]
where \(k=26\)s/L represents our calibrated flow constant. This reduces computational overhead remarkablely compared to sensor-based systems while maintaining sufficient accuracy for target applications.

\section{System Architecture and Design}

\subsection{System Overview}
The low-cost fuel dispensing system integrates three core subsystems to deliver an efficient, user-friendly solution for fuel dispensing. These subsystems collaborate seamlessly: the keypad captures user inputs, the microcontroller processes these inputs to compute fuel volume and control system behavior, and the output layer—comprising an LCD and a motor—provides feedback and executes the dispensing. Specifically, the 4x4 keypad sends user-entered fuel amounts and commands to the STM32F103C8T6 microcontroller, which calculates the motor runtime based on a calibrated flow rate and updates the 16x4 LCD with real-time status information. The motor, in turn, dispenses the precise fuel volume under microcontroller supervision. Figure 1 illustrates this workflow, depicting the data flow from input to processing to output, while Figure 2 details the electrical interconnections that enable this interaction.

\begin{figure}[H]
\centering
\includegraphics[width=0.9\linewidth]{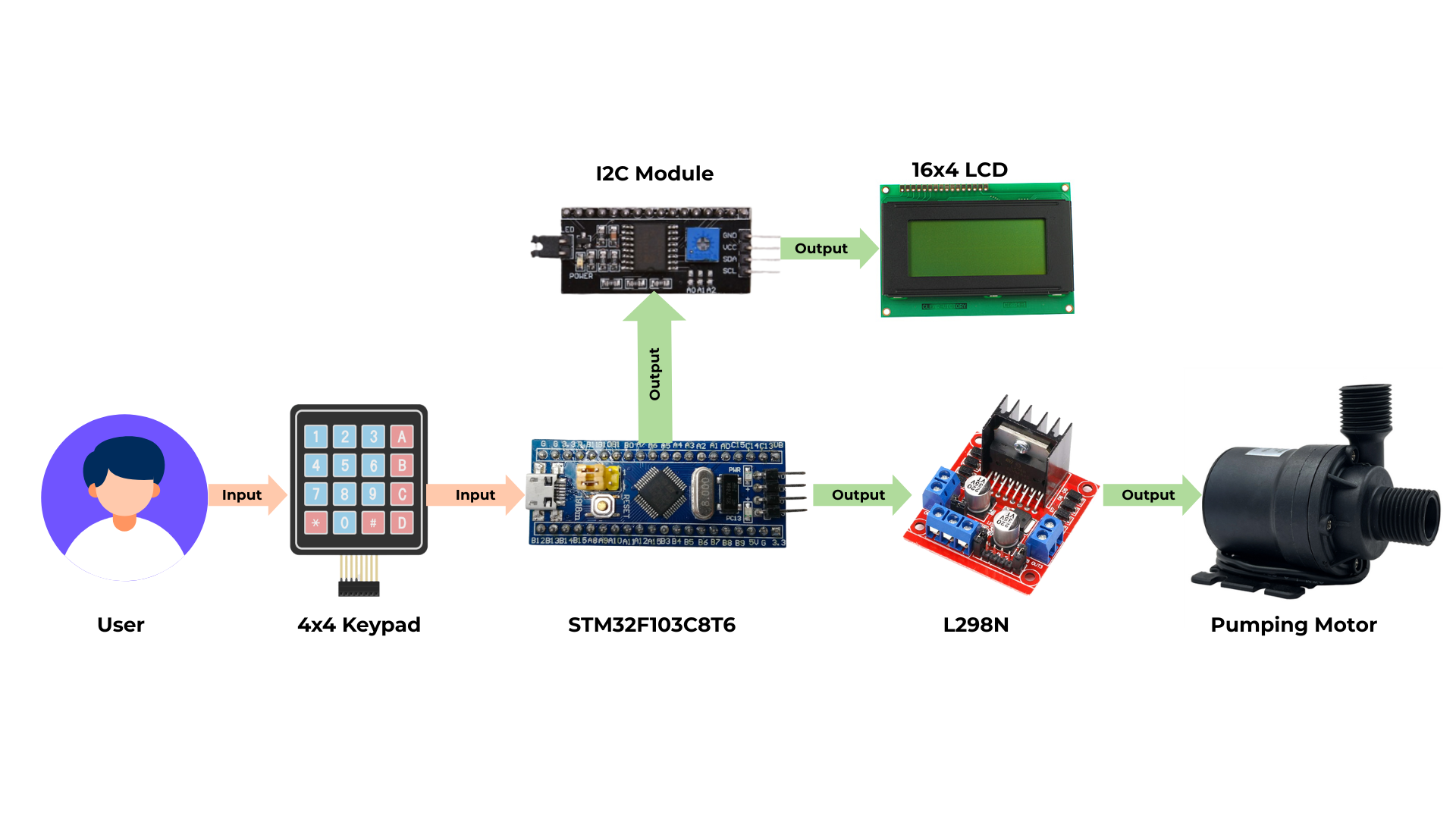}
\caption{Data work flow diagram of the fuel dispensing system.}
\label{fig:flowchart}
\end{figure}

\subsection{Hardware Architecture}
The system's hardware is designed for reliability and cost-efficiency, with each component carefully selected to fulfill a specific role in the dispensing process. Below are the key components and their detailed specifications:

\begin{enumerate}
    \item \textbf{STM32F103C8T6 Microcontroller:}
    \begin{itemize}
        \item \textbf{Processor:} ARM Cortex-M3 core running at 72MHz, delivering sufficient computational power for real-time input processing, motor control, and display updates. [10],[11]
        \item \textbf{Memory:} 128KB Flash for storing the embedded firmware and 20KB RAM for handling runtime variables and temporary data buffers.
        \item \textbf{Interfaces:} I2C (for LCD communication), GPIO (for keypad scanning and motor control), and timers for precise motor runtime calculations.
        \item \textbf{Power:} Operates at 3.3V with a maximum current draw of 50mA, optimized for energy efficiency in battery-powered applications.
        \item \textbf{Features:} Includes an internal oscillator and multiple interrupt channels, enhancing responsiveness to user inputs and system events.
    \end{itemize}
    The microcontroller acts as the central hub, orchestrating all subsystem interactions with high reliability and low latency. [12]

    \item \textbf{Motor Control Subsystem: L298N Dual H-Bridge Motor Driver:}
    \begin{itemize}
        \item \textbf{Design:} A robust dual H-bridge motor driver IC capable of controlling the direction and speed of two DC motors independently, widely used in embedded motor applications. [13]
        \item \textbf{Voltage Support:} Operates with input voltages from 5V to 35V and supports output current up to 2A per channel, suitable for driving a 12V DC pump motor used in our prototype.
        \item \textbf{Connection:} Connected to the STM32 GPIO pins PA8 (IN3) and PA9 (IN4), enabling directional control via logic high/low signals.
        \item \textbf{Operation:} When IN3 is HIGH and IN4 is LOW, the pump motor runs forward; both LOW signals stop the motor—ensuring safe, predictable control using simple digital logic.
        \item \textbf{Features:} Includes built-in protection diodes for back EMF suppression and onboard heat sinks for thermal management during continuous operation.
    \end{itemize}
    The L298N acts as the power interface between the microcontroller and the mechanical pump system, ensuring reliable and efficient fuel dispensing under varying electrical loads.

    \item \textbf{Input Subsystem: 4x4 Matrix Keypad:}
    \begin{itemize}
        \item \textbf{Design:}  A 16-button tactile keypad in a 4x4 matrix layout, minimizing pin usage via row-column scanning.
        \item \textbf{Connections:} Columns wired to PA0-PA3 and rows to PA4-PA7 on the microcontroller, leveraging GPIO pull-up resistors for debouncing.
        \item \textbf{Functions:} Features digits (0-9) for entering fuel amounts, a decimal point for precise inputs (e.g., 2.5 liters), and control keys: backspace (B) to correct entries, clear (*) to reset, confirm (A) to initiate dispensing, and stop (\#) to halt operations.
        \item \textbf{Durability:} Rated for 100,000 presses per key, ensuring longevity in frequent-use scenarios.
    \end{itemize}
    The keypad offers an intuitive interface, enabling users to interact with the system effortlessly. [14]
\\
    \item \textbf{Output Subsystems:}
    \begin{itemize}
        \item \textbf{16x4 LCD Display:}
        
        \begin{itemize}
            \item \textbf{Interface:} Connected via I2C (SDA on PB7, SCL on PB6), reducing pin count and simplifying wiring.

            \item \textbf{Function:} Displays prompts (e.g., "Enter amount"), user inputs, system status (e.g., "Dispensing"), and dispensed volume, with a 16-character width and 4-line capacity for detailed feedback. [15]
            \item \textbf{Contrast:} Adjustable via a potentiometer, ensuring readability in varying light conditions.[16]
        \end{itemize}
    \end{itemize}
The LCD enhances usability by providing clear, immediate visual feedback throughout the dispensing process.\\

\end{enumerate}

\begin{figure}[H]
\centering
\includegraphics[width=0.8\linewidth]{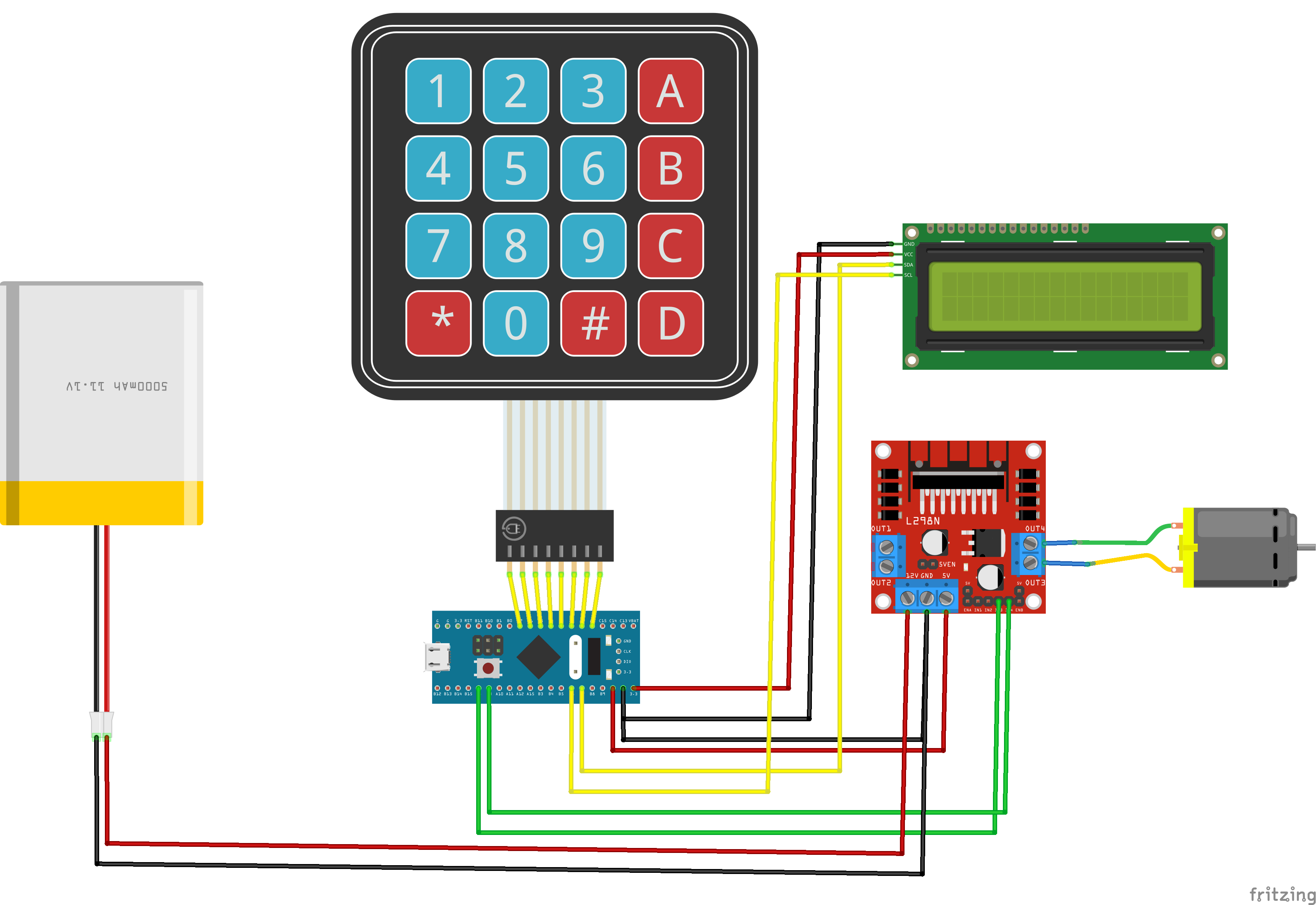}
\caption{Circuit diagram of the system.}
\label{fig:flowchart}
\end{figure}

\newpage

\section{Control Algorithm and Flowchart}

The fuel dispensing system's control logic is structured into five primary phases: Initialization, User Input, Motor Operation, Idle State, and the Main Loop. Each phase contributes to accurate fuel dispensing through real-time input handling and motor control. The algorithm ensures user-friendly interaction via the keypad and provides feedback using an LCD. 

\subsection{Algorithm Steps}

\begin{algorithm} [H]
\caption{Main Control Logic for Fuel Dispenser Prototype}
\begin{algorithmic}[1]
\STATE Initialize system components
\STATE Initialize LCD via I2C
\STATE Initialize 4×4 Keypad
\STATE Configure motor control pins and keypad I/O
\STATE Display ``Enter Amount'' on LCD
\STATE $input\_buffer \leftarrow$ empty string
\STATE $motor\_running \leftarrow$ FALSE
\WHILE{system is powered}
    \IF{$motor\_running = \text{TRUE}$}
        \IF{stop key (\#) is pressed}
            \STATE Stop motor immediately
            \STATE $motor\_running \leftarrow$ FALSE
            \STATE Reset $input\_buffer$
            \STATE Display ``Enter Amount'' on LCD
        \ELSIF{motor duration elapsed}
            \STATE Stop motor normally
            \STATE $motor\_running \leftarrow$ FALSE
            \STATE Reset $input\_buffer$
            \STATE Display ``Enter Amount'' on LCD
        \ELSE
            \STATE Update LCD with current dispensing status
        \ENDIF
    \ENDIF
    \STATE $key \leftarrow$ Poll keypad
    \IF{$key \neq \text{null}$}
        \STATE \textbf{switch} (key)
        \STATE \quad \textbf{case} digit/decimal:
        \STATE \qquad Append key to $input\_buffer$
        \STATE \qquad Display $input\_buffer$ on LCD
        \STATE \quad \textbf{case} backspace:
        \STATE \qquad Remove last character from $input\_buffer$
        \STATE \qquad Update LCD accordingly
        \STATE \quad \textbf{case} clear:
        \STATE \qquad Reset $input\_buffer$
        \STATE \qquad Display ``Enter Amount'' on LCD
        \STATE \quad \textbf{case} confirm:
        
        \STATE \textbf{end switch}
    \ENDIF
\ENDWHILE
\end{algorithmic}
\end{algorithm}

\subsection{Flowchart}

\begin{figure}[H]
\centering
\includegraphics[width=0.8\linewidth]{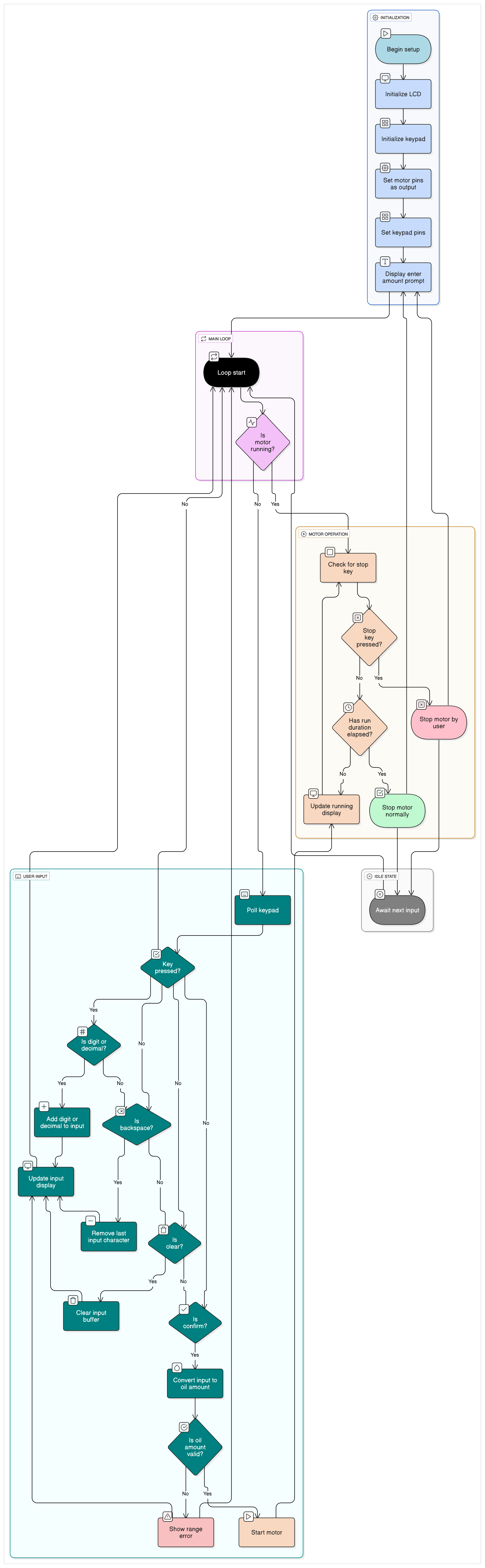}
\caption{Control flowchart of the fuel dispensing system.}
\label{fig:flowchart}
\end{figure}

The flowchart above (Figure 3) visually represents the logical sequence of operations, starting from system initialization to input validation, motor control, and idle state. It illustrates both the sequential and conditional flows of control based on user actions and internal system states.

\section{System State-Space Equation}

The fuel dispensing system is modeled as a first-order linear time-invariant (LTI) system where the volume of fuel dispensed is directly proportional to the runtime of the DC pump motor. Based on experimental calibration, the motor takes approximately 26 seconds to dispense 1 liter of fuel under standard operating conditions. This fixed flow rate allows us to define a state-space representation of the system.

Let:
\begin{itemize}
    \item \( x(t) \) be the volume of fuel dispensed at time \( t \) (in liters),
    \item \( u(t) \in \{0,1\} \) be the motor control input, where 1 indicates the motor is running and 0 indicates it is off,
    \item \( r = \frac{1}{26} \approx 0.0385 \) L/s be the rate at which the fuel is dispensed when the motor is on.
\end{itemize}

\subsection*{State Equation}
The continuous-time state-space equation of the system is given by:
\[
\dot{x}(t) = r \cdot u(t)
\]
\[
\dot{x}(t) = \frac{1}{26} u(t)
\]

\subsection*{Output Equation}
Assuming the output \( y(t) \) is the same as the state variable (volume of fuel dispensed), we define:
\[
y(t) = x(t)
\]

\subsection*{Matrix Form}
In [17] standard state-space matrix form, the system can be represented as:
\[
\dot{x}(t) = A x(t) + B u(t)
\]
\[
y(t) = C x(t) + D u(t)
\]
where:
\[
A = 0,\quad B = \frac{1}{26},\quad C = 1,\quad D = 0
\]

\subsection*{Interpretation}
This model represents an integrator system with constant gain. When the user inputs a desired fuel volume, the controller calculates the corresponding motor runtime using the inverse of the flow rate (i.e., \(26 \, \text{seconds/liter}\)). The STM32 microcontroller ensures that the motor runs for the exact duration to dispense the correct amount of fuel.

\section{Results and Discussion}

\subsection{Working Prototype}
The prototype of the low-cost fuel dispensing system was evaluated using water as a safe and accessible substitute for fuel. In Figure 4, the full working prototype is demonstrated. The following key findings emerged from the testing process:

\begin{itemize}
    \item \textbf{Dispensing Accuracy:} The system dispensed water volumes with significant precision, indicating reliable performance under controlled conditions.
    \item \textbf{User Interface:} The 4x4 keypad and 16x2 LCD screen provided an intuitive experience, enabling users to operate the system effectively without training.
    \item \textbf{System Responsiveness:} The STM32 microcontroller and L298N motor driver ensured quick and consistent pump activation, delivering a seamless operation.
    \item \textbf{Limitations:} Performance with actual fuels (e.g., gasoline, diesel) may differ due to variations in density and viscosity, necessitating further calibration.
\end{itemize}

\begin{figure}[H]
\centering
\includegraphics[width=0.9\linewidth]{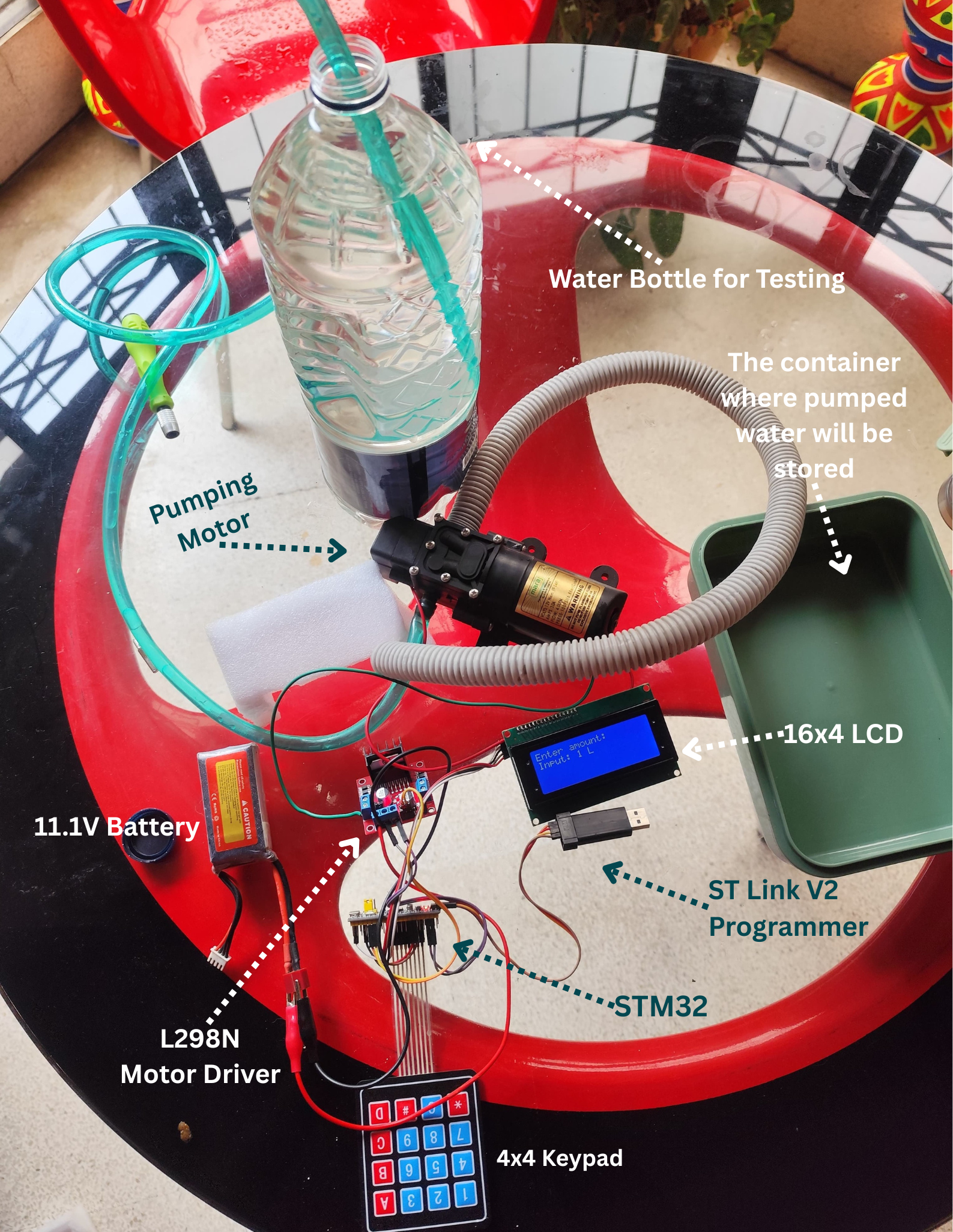}
\caption{Working Prototype of Low Cost Fuel Dispenser.}
\label{fig:flowchart}
\end{figure}

\subsection{Future Scalability}
The successful testing of the prototype highlights its potential as an affordable and portable fuel dispensing solution, leveraging embedded systems technology. The following points address its implications and future directions:

\begin{itemize}
    \item \textbf{Feasibility:} The system’s cost-effectiveness and modularity make it viable for small-scale or remote applications where conventional dispensers are impractical.
    \item \textbf{Future Enhancements:}
        \begin{itemize}
            \item Integration of flow sensors to improve real-time volume accuracy across fuel types.
            \item Addition of GSM modules for remote monitoring and management.
            \item Implementation of RFID cards for secure user authentication.
            \item Incorporation of payment systems to enable commercial deployment.
        \end{itemize}
    \item \textbf{Scalability:}
        \begin{itemize}
            \item The design supports expansion to multiple dispensing units for larger operations.
            \item It can be adapted into broader fuel management systems, enhancing accessibility in underserved areas.
        \end{itemize}
    \item \textbf{Next Steps:} Future efforts will focus on calibrating the system for specific fuels and testing scalability in real-world settings.
\end{itemize}

\section{Conclusion}
This paper presented a cost-effective, ARM-based fuel dispensing system utilizing the STM32F103C8T6 platform, L298N motor driver, and a user-friendly keypad interface. The proposed design achieves accurate volume-based dispensing without the need for expensive sensors or wireless connectivity. Our prototype demonstrates how embedded systems can simplify tasks in resource-constrained environments, offering a scalable solution for local fuel management needs.

\vspace{12pt}

\end{document}